\documentclass[12pt, letterpaper, superscriptaddress, final]{iopart}

\expandafter\let\csname equation*\endcsname\relax
\expandafter\let\csname endequation*\endcsname\relax

\usepackage{amsmath,amssymb}
\usepackage{graphicx}
\usepackage{dcolumn}
\usepackage{bm}
\usepackage{hyperref}
\usepackage{url}
\usepackage{braket}
\usepackage{color}

\usepackage{enumitem}

\usepackage{titlecaps}
\Addlcwords{the of and to}

\usepackage{longtable}
\usepackage{makecell}
\usepackage{etoolbox}

\makeatletter
\newcommand{\mainmatter}{%
  \setcounter{footnote}{0}%
  \patchcmd{\@makefntext}{\fnsymbol}{\arabic}{}{}%
  \patchcmd{\@thefnmark}{\fnsymbol}{\arabic}{}{}%
  \def\@makefnmark{\textsuperscript{\arabic{footnote}}}%
}
\makeatother

\begin{document}

\title[Wigner time delays and Goos-H\"{a}nchen shifts of 2D quantum vortices]{Wigner time delays and Goos-H\"{a}nchen shifts of 2D quantum vortices scattered by potential barriers}

\author{Maxim Mazanov$^1$ and Konstantin Y. Bliokh$^2$}
\address{$^1$School of Physics and Engineering, ITMO University, St. Petersburg, Russia}
\address{$^2$Theoretical Quantum Physics Laboratory, RIKEN Cluster for Pioneering Research, Wako-shi, Saitama 351-0198, Japan}

\begin{abstract}
We consider reflection and transmission of 2D quantum wavepackets with phase vortices (also known in optics as spatiotemporal vortex pulses) at potential step-like, delta-function, and rectangular barriers. The presence of a vortex significantly modifies the Wigner time delays and Goos-H\"{a}nchen shifts, previously studied for Gaussian-like wavepackets. In particular, the scattered wavepackets undergo non-zero time delays and lateral shifts even for purely real scattering coefficients, when the standard Wigner and Artmann formulae vanish. We derive analytical expressions for the vortex-induced times delays and spatial shifts of 2D vortices and verify these with numerical calculations of the Schr\"{o}dinger equation. The time delays and shifts are resonantly enhanced in the vicinity of the critical-angle incidence for a step-like potential and near transmission resonances for a rectangular barrier.  
\end{abstract}

%
%
%
%
%



\section{Introduction}

Interference of waves in complex fields results in numerous nontrivial phenomena, such as phase singularities (vortices) and other topological wave structures \cite{Nye1974,Berry1981,Rubinsztein2016}, superoscillations and quantum weak values \cite{Aharonov1988,Berry2019,Dressel2014}, time delays and superluminal propagation \cite{Chiao1997,Carvalho2002,Winful2006}, beam shifts and Hall effects \cite{Bliokh2013,Gotte2012,Toppel2013,Bliokh2015NP,Xiao2017}, etc. In some cases, these phenomena exhibit remarkable interplay with each other \cite{Hosten2008,Berry2009,Bliokh2013NJP_II,Asano2016}. Moreover, because of their universal wave nature, these effects appear equally in classical (e.g., optical or acoustic) and quantum (condensed-matter or free-particle) systems \cite{Dragoman}. 

In this work, we examine a problem with remarkable interplay of the Wigner time delays \cite{Wigner1954}, Goos-H\"anchen (GH) shifts \cite{Goos1947,Artmann1948}, and phase vortices \cite{Nye1974,Berry1981} in a basic quantum system. Namely, we consider reflection and transmission of a 2D vortex wavepacket at various types of planar potential barriers within the Schr\"{o}dinger equation. 

Reflection and transmission of vortex beams, with the accompanying beam shifts, was thoroughly studied in optics \cite{Fedoseyev2001,Dasgupta2006,Fedoseyev2008,Bliokh2009,Merano2010,Dennis2012}. The presence of a vortex and the corresponding orbital angular momentum (OAM) dramatically modifies the beam shifts as compared to the case of Gaussian-like wavepackets. However, such problems were considered for the 3D setup where the vortex line and the corresponding OAM are aligned with the beam propagation direction (i.e., the beam carries a {\it screw} phase dislocation \cite{Nye1974,Berry1981}). Recently, there was considerable interest in {\it spatiotemporal} vortex pulses carrying the vortex line and OAM in the direction orthogonal to the propagation \cite{Bliokh2012,Jhajj2016,Hancock2019,Chong2020,Bliokh2021,Zang2022,Junyi2022} (i.e., the pulse carries an {\it edge} phase dislocation \cite{Nye1974,Berry1981}). By examining reflection and refraction of such optical pulses at planar dielectric interfaces, we recently found that the spatiotemporal vortex and transverse OAM cause a novel effect of time delay (longitudinal pulse shift) and enables either subluminal or superluminal propagation of the reflected/transmitted pulses \cite{Mazanov2021}. Remarkably, this vortex-induced time delay cannot be calculated from the standard Wigner formula \cite{Wigner1954,Chiao1997,Carvalho2002,Winful2006} which is valid only for Gaussian-like pulses without singularities and requires a temporal dispersion of the interface properties.

An important difference between the geometries of the longitudinal and transverse OAM/vortices is that the latter can naturally exist in planar 2D systems. In particular, 2D vortices carrying orthogonal OAM naturally appear in quantum condensed-matter systems, such as superfluids, BEC, quantum-Hall systems, 2D electron gas, and ferromagnets \cite{Huebener,Fetter2009,Nagaosa2013,Thouless1996,Stone1996}. Furthermore, the Wigner time delays and superluminal tunneling were originally considered for quantum wavepackets \cite{Wigner1954,Chiao1997,Carvalho2002,Winful2006}. Therefore, the problem of vortex-induced time delays and shifts naturally arises for scattering (reflection or refraction) of a 2D quantum vortex at a potential barrier. 

Here we examine this problem by considering the propagation and scattering of a 2D Laguerre-Gaussian-type vortex wavepacket within the Schr\"{o}dinger equation with planar potential obstacles of different types: a step, a delta-function, and a rectangular barrier. We calculate the GH shifts and Wigner time delays of the reflected and transmitted wavepackets both in space-time (real or `linear' shifts) and momentum-energy (imaginary or `angular' shifts). We find novel vortex-induced contributions to the GH shifts and Wigner time delays of the scattered wavepackets. Most importantly, the vortex-induced time delays and lateral shifts appear even for purely real scattering coefficients where the standard Wigner time delay and GH effect vanish. These findings could be relevant to nontrivial transport properties of vortex states in 2D quantum systems. 


\section{Scattering of plane waves and Gaussian-like wavepackets}

\subsection{Schr\"{o}dinger equation. Reflection and refraction of a plane wave}

\begin{figure}[!t]
\begin{center}
\includegraphics[width=0.85\linewidth]{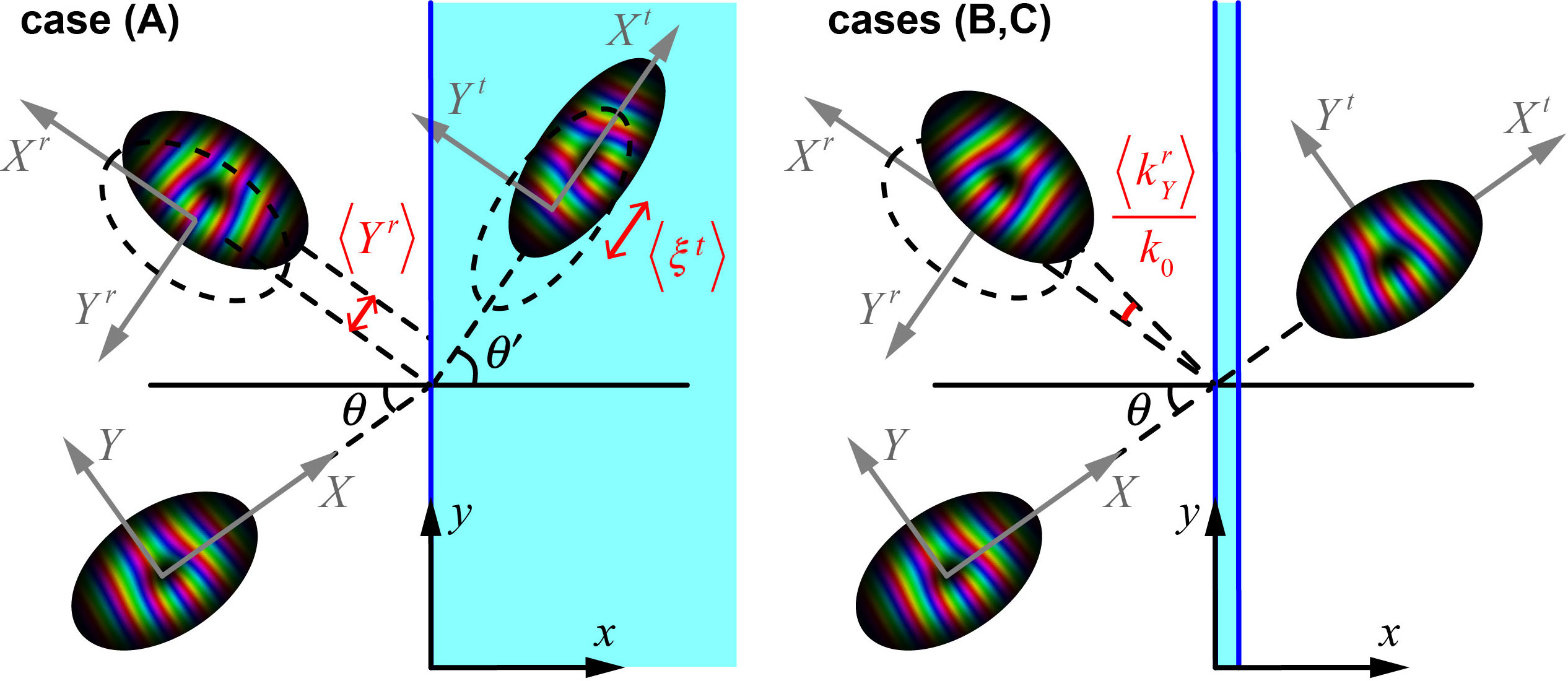}
\caption{Reflection and refraction of a wavepacket (here, with a vortex of topological charge $\ell=1$) at potential barriers. The wavepackets are schematically shown using brightness proportional to the probability density $|\psi |^2$ and colour-coded phase ${\rm Arg}(\psi)$ \cite{Thaller_book}.
The case (A) corresponds to a step potential, whereas the cases (B) and (C) correspond to delta-function and thin rectangular barriers, respectively (shown schematically on the same plot). Examples of the linear GH and angular GH shifts, $\left\langle Y \right\rangle$  and $\left\langle k_Y \right\rangle$, as well as of the longitudinal (Wigner time delay) shift $\left\langle \xi \right\rangle$ are shown schematically for some of the reflected and transmitted wavepackets.
\label{Fig1}}
\end{center}
\end{figure}

Throughout this work we deal with the 2D Schr\"{o}dinger equation using the units $\hbar = m =1$:
\begin{equation}
\label{eq1}
i \frac{\partial}{\partial t}\psi({\bf r},t) = \left[ -\frac{{\bm\nabla}^2}{2} + V(x) \right]\! \psi({\bf r},t)\,,
\end{equation}
where $\psi$ is the wavefunction, ${\bf r} = (x,y)$, ${\bm\nabla}=\partial/\partial{\bf r}$, and $V(x)$ is the potential. We consider three examples of potential barriers:
\begin{enumerate}[label=(\Alph*)]
\item Step potential $V(x) = V_0\, \Theta(x)$, where $\Theta(x)$ is the Heaviside step-function;
\item Delta-function barrier $V(x) = W_0\, \delta(x)$, where $\delta(x)$ is the Dirac delta-function;
\item Rectangular barrier $V(x) = V_0 \left[ \Theta(x) - \Theta(x-a) \right]$.
\end{enumerate}
We will consider reflection and transmission of plane waves and wavepackets incident from the $x<0$ half-plane at the angle $\theta$ with respect to the $x$-axis, see Fig.~\ref{Fig1}. 

\begin{figure}[!t]
\begin{center}
\includegraphics[width=0.75\linewidth]{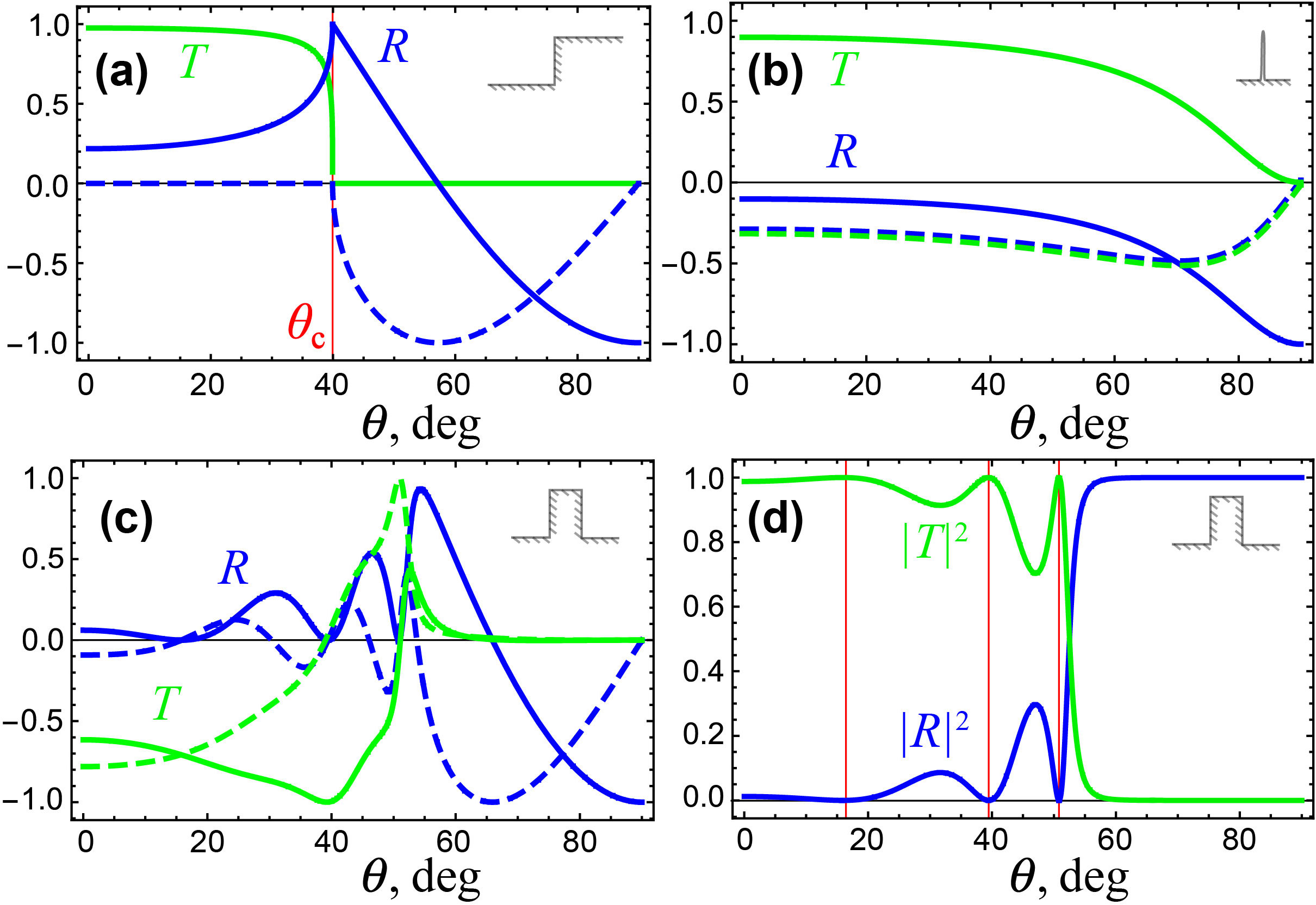}
\caption{The plane-wave reflection and transmission amplitudes (\ref{eq2})--(\ref{eq4}) versus the angle of incidence, $\theta$, for a step potential with $E/V_0 = 1.7$ (a); a delta-function potential with $k/W_0 = 3$ (b); and rectangular potential with $E/V_0 = 3$ and $k a = 5$ (c). The real and imaginary parts of the amplitudes are depicted with solid and dashed curves, respectively. The panel (d) shows the energy transmission and reflection coefficients, $|T|^2$ and $|R|^2$, for the amplitudes from (c).
\label{Fig2}}
\end{center}
\end{figure}

Consider first a single incident plane wave, $\psi \propto \exp \left( -i\,\omega\, t + i\,{\bf k}\cdot {\bf r} \right)$, $x \leq 0$, with energy (frequency) $E=\omega$, momentum (wave vector) ${\bf p}={\bf k} \equiv (k_x,k_y) = k (\cos\theta, \sin\theta)$, and dispersion $\omega = k^2/2$ following from Eq.~(\ref{eq1}). In the step-potential case (A), the reflected and transmitted waves, $\psi^{r} \propto R \exp \left( -i\,\omega\, t + i\,{\bf k}^r \cdot {\bf r} \right)$, $x\leq 0$, and $\psi^{t} \propto T \exp \left( -i\,\omega\, t + i\,{\bf k}^t \cdot {\bf r} \right)$, $x\geq 0$, have the same energies and wavevectors ${\bf k}^{r} = k (-\cos\theta, \sin\theta)$ and ${\bf k}^{t} = k' (\cos\theta', \sin\theta')$, where $k' = k \sqrt{1-V_0/E}$ and $\sin\theta' = \sin\theta / \sqrt{1-V_0/E}$ \cite{Dragoman}. Applying the appropriate boundary conditions, i.e., the continuity of the total wavefunction $\psi$ and its normal derivative $\partial\psi/\partial x$ at $x=0$, we obtain the scattering (reflection and transmission) amplitudes [see Fig.~\ref{Fig2}(a)]: 
\begin{equation}
R = \frac{k_x - k'_x}{k_x + k'_x}\,, \quad
T = \frac{2\sqrt{k_x k'_x}}{k_x + k'_x}\,,
\label{eq2} 
\end{equation}
where $k'_x = k'\cos\theta'$. Note that the reflected and transmitted waves exist and acquire real amplitudes (\ref{eq2}) in the regime of {\it partial reflection/transmission}, $\theta < \theta_c = \arccos \sqrt{V_0/E}$, while only the reflected propagating wave is generated in the regime of {\it total reflection} $\theta > \theta_c$, where one should use $k'_x = i \sqrt{2(V_0-E\cos^2\theta)}$, $T=0$, and $R$ becomes complex. 

In the delta-potential case (B), the incident wave is partially reflected/transmitted for all angles $\theta \in (0,\pi/2)$, whereas the transmitted wave has the same wavevector as the incident one, ${\bf k}^t ={\bf k}$. The boundary conditions are: the continuity of the total wavefunction at $x=0$ and relation $\left.\dfrac{\partial\psi}{\partial x} \right|_{x=+0} - \left.\dfrac{\partial\psi}{\partial x} \right|_{x=-0} = 2W_0$ which follows from the integration of the Schr\"{o}dinger equation (\ref{eq1}) across the delta-barrier, $x\in (-0,+0)$. This results in the complex reflection and transmission amplitudes [see Fig.~\ref{Fig2}(b)]
\begin{equation}
R = - \frac{i W_0}{k_x + i W_0}\,, \quad
T = \frac{k_x}{k_x + i W_0}\,. 
\label{eq3} 
\end{equation}

Finally, in the case (C) of a finite rectangular barrier, consisting of two potential-step interfaces, the complex reflection and transmission amplitudes read [see Fig.~\ref{Fig2}(c)]
\begin{equation}
R = \frac{\left(k_x^2-k'^2_x\right) \sin (k'_x a)}{\left(k_x^2 + k'^2_x\right) \sin (k'_x a)+2 i k_x k'_x \cos ( k'_x a)}, ~~
T = \frac{2i k_x k'_x e^{-i k_x a}}{\left(k_x^2 + k'^2_x\right) \sin (k'_x a)+2 i k_x k'_x \cos ( k'_x a)},
\label{eq4} 
\end{equation}
where $k'_x = k'\cos\theta'$ is the normal wavevector component inside the barrier. Note that the scattering coefficients (\ref{eq2})--(\ref{eq4}) satisfy the conservation law $|R|^2 + |T|^2 =1$, and scattering by a rectangular barrier (C) exhibits resonances with total transmission $|T|=1$, $|R|=0$, Fig.~\ref{Fig2}(d).

\subsection{Goos-H\"{a}nchen shifts and Wigner time delays for Gaussian-like wavepackets}

We now consider reflection and transmission of a Gaussian wavepacket at the potential barriers of types (A)--(C). The incident wavepacket can be characterized by its Fourier spectrum (momentum representation) 
\begin{equation}
\tilde{\psi}(\mathbf{k}) \propto \exp\! \left\{ -\frac{\Delta^{2}}{4}\left[\gamma^{2}\!\left(k_{X}-k_{0}\right)^{2}+k_{Y}^{2}\right] \right\} .
\label{eq5} 
\end{equation}
Here and hereafter we use the coordinate frame $(X,Y)$ with the $X$-axis oriented along the propagation of the incident wavepacket, i.e., rotated by the angle $\theta$ with respect to the $(x,y)$ frame (see Fig.~\ref{Fig1}), $k_0$ is the central wavevector in the wavepacket, $\Delta$ is the $Y$-width of the packet, whereas $\gamma\Delta$ is its $X$-length. We assume the wavepacket (\ref{eq5}) to be paraxial and large as compared with the wavelength (i.e., semiclassical): $\{k_0\Delta, \gamma k_0\Delta\} \gg 1$. In the case (C) of a rectangular barrier we also assume that the wavepacket is much larger than the barrier width: $\{\Delta, \gamma\Delta \} \gg a$; this is to avoid numerous maxima in the reflected/transmitted signal due to the multiple reflections of a short pulse inside the barrier \cite{Winful2006}. 

\begin{figure}[!t]
\begin{center}
\includegraphics[width=0.65\linewidth]{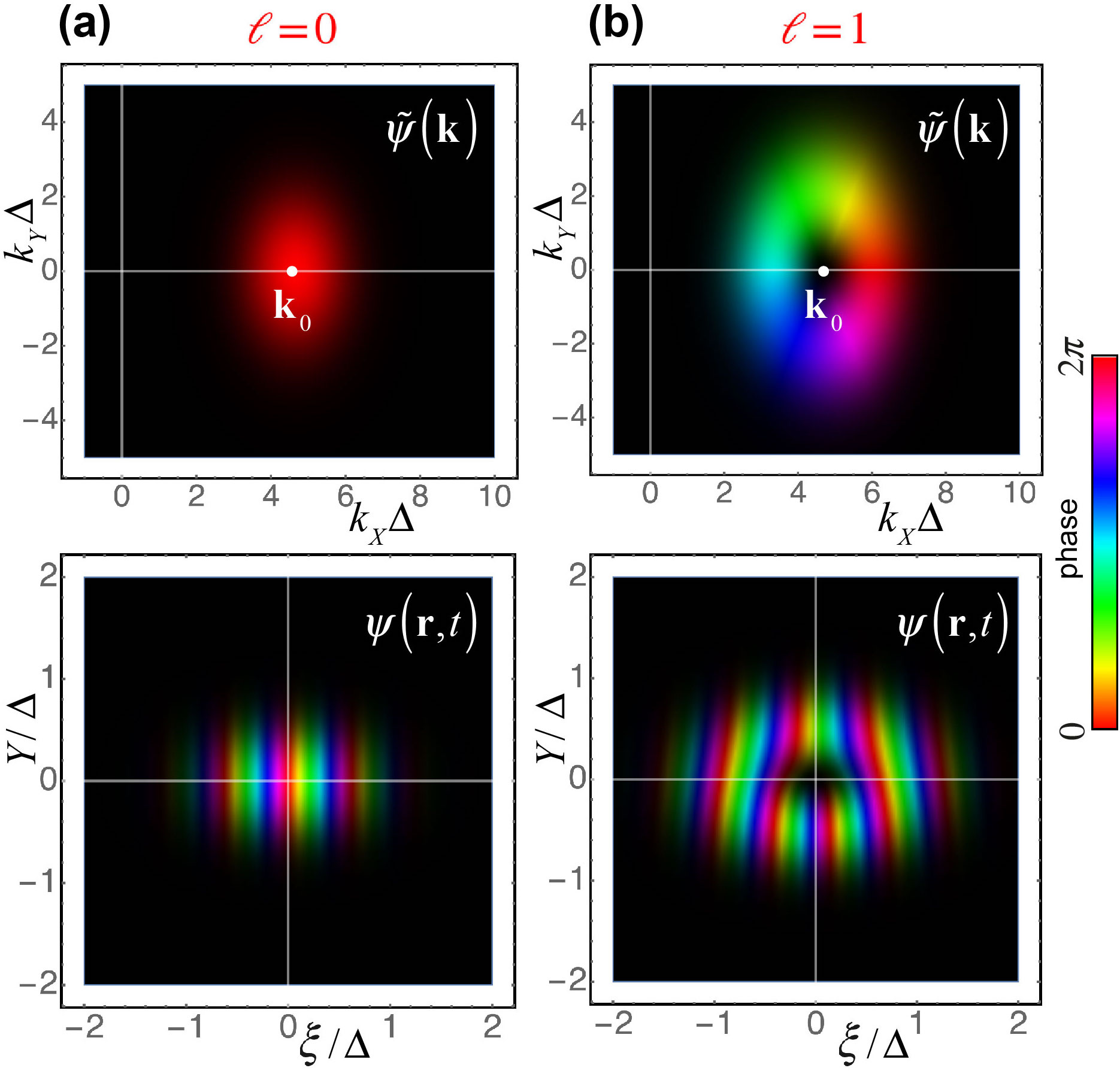}
\caption{Phase-intensity distributions of the momentum-space (upper panels) and real-space (lower panels) wavefunctions of Gaussian ($\ell=0$) and Laguerre-Gaussian (with the vortex $\ell=1$ here) wavepackets, Eqs.~(\ref{eq10}) and (\ref{eq11}). The parameters are $k_0\Delta = 0.7$ and $\gamma = 1.5$. The brightness is proportional to the probability density $|\psi |^2$, whereas the colour indicates the phase ${\rm Arg}(\psi)$ \cite{Thaller_book}.
\label{Fig3}}
\end{center}
\end{figure}

The real-space (coordinate) representation of the Gaussian wavepacket (\ref{eq5}) is given by the Fourier integral
\begin{equation}
\psi \left( {{\bf{r}},t} \right) \propto \iint {\tilde \psi \left( {\bf{k}} \right){e^{i{\bf{k}} \cdot {\bf{r}} - i\omega \left( {\bf{k}} \right)t}}} d{k_X}d{k_Y} \propto \exp\! \left[ { - \frac{\left( {{\gamma ^{ - 2}}{\xi ^2} + {Y^2}} \right)}{{{\Delta ^2}}} + i{k_0}X - i \omega_0 t } \right]\! ,
\label{eq6} 
\end{equation}
where $\omega \left( {\bf{k}} \right) = k^2/2$, $\xi = X - v_g\, t$ is the wavepacket-accompanying coordinate with the group velocity $v_g = k_0$, and we neglected diffraction effects in the paraxial approximation. An example of the momentum-space and real-space wavefunctions (\ref{eq5}) and (\ref{eq6}) is shown in Fig.~\ref{Fig3}(a). 

Spatial dispersion of the scattering coefficients (\ref{eq2})--(\ref{eq4}), i.e., their dependence on the angle of incidence, $\theta$, causes the GH shifts of the scattered wavepackets \cite{Goos1947,Artmann1948,Bliokh2013}. The linear (position) and angular (direction) GH shifts can be characterized by the mean transverse position $\langle Y \rangle$ and transverse momentum $\langle k_Y \rangle$ (with respect to the expected geometrical-optics or semiclassical trajectory) of the scattered wavepackets. The generalized Artmann formula \cite{Artmann1948,Bliokh2013} yields
\begin{eqnarray}
\label{eq7} 
&& \langle Y^r \rangle_0 = \frac{1}{k_0} {\rm Im}\! \left( \frac{\partial \ln R}{\partial \theta}  \right)
\!, \quad
\langle Y^t \rangle_0 = - \frac{\cos\theta^t}{k_0\cos\theta} {\rm Im}\! \left( \frac{\partial \ln T}{\partial \theta}  \right)\!,
\\ 
 \label{eq8} 
&& \langle k_Y^r \rangle_0  = - \frac{2}{k_0 \Delta^2} {\rm Re}\! \left( \frac{\partial \ln R}{\partial \theta}  \right)\! , 
\quad
\langle k_Y^t \rangle_0  = \frac{2\cos\theta}{k_0 \Delta^2 \cos\theta^t} {\rm Re}\! \left( \frac{\partial \ln T}{\partial \theta}  \right)\!.
\end{eqnarray}
Here and hereafter the coordinates and wavevectors of the reflected and transmitted wavepackets are written in the coordinate frames $(X^{r,t},Y^{r,t})$ accompanying the reflected and refracted wavepackets \cite{Bliokh2013}, the subscript ``0'' indicates that we deal with a Gaussian (no-vortex) wavepacket, $\theta^t=\theta'$ in the case of step potential (A), and $\theta^t=\theta$ in the cases (B) and (C). The coordinate frames and schematics of some of the GH shifts are shown in Fig.~\ref{Fig1}. The linear shifts are non-diffractive effects appearing right after the scattering event, while the angular shifts accumulate additional spatial shifts proportional to the propagation distances $X^{r,t}$ \cite{Bliokh2013,Chan1985,Aiello2008,Merano2009}. The GH shifts are well studied in optics and also in quantum systems \cite{Beenakker2009,Haan2010,Wu2011,Chen2013}, although, to the best of our knowledge, the angular GH effect has never been discussed for quantum systems.

In analogy to the GH effect, the temporal dispersion of the scattering coefficients (\ref{eq2})--(\ref{eq4}), i.e., their dependence on the energy of the incident wave, $E$, results in the Wigner time delays of the scattered wavepackets \cite{Wigner1954,Chiao1997,Carvalho2002,Winful2006,Asano2016}.
\begin{eqnarray}
\label{eq9} 
&& \langle \tau^r \rangle_0 = {\rm Im}\! \left( \frac{\partial \ln R}{\partial E}  \right)
\!, \quad
\langle \tau^t \rangle_0 =  {\rm Im}\! \left( \frac{\partial \ln T}{\partial E}  \right)\!,
 \\ 
 \label{eq10} 
&& \langle \epsilon^r \rangle_0  = \frac{2 k_0^2}{\gamma^2\Delta^2} {\rm Re}\! \left( \frac{\partial \ln R}{\partial E}  \right)\! , 
\quad
\langle \epsilon^t \rangle_0  = \frac{2 k_0^2}{\gamma^2 \Delta^2} {\rm Re}\! \left( \frac{\partial \ln T}{\partial E}  \right)\!.
\end{eqnarray}
Here $\langle \tau \rangle$ denotes the time delay of the scattered wavepacket as compared to its expected arrival time, whereas $\langle \epsilon \rangle$ is the shift of its mean energy/frequency with respect to the central energy of the incident wavepacket, $E_0=\omega_0 = k_0^2/2$. These are temporal counterparts of the linear and angular GH shifts, so that the energy shift accumulates an additional time delay proportional to the propagation time of the wavepacket. The Wigner time delays (\ref{eq9}) and (\ref{eq10}) can be equally regarded as spatial and angular shifts in the longitudinal wavepacket-accompanying coordinate $\langle \xi^{r,t} \rangle = - v^{r,t}_g \langle \tau^{r,t} \rangle$ and the corresponding wavevector component $\langle \delta k^{r,t}_X \rangle = (v^{r,t}_g)^{-1} \langle \epsilon^{r,t} \rangle$, where $\xi^{r,t} = X^{r,t} - v^{r,t}_g t$, $\delta k^{r,t}_X = k^{r,t}_X - k^{r,t}_0$, $v^t_g = k^t_0 = k'_0$ in the step-potential case (A) and $v^{r,t}_g = k^{r,t}_0 = k_0$ in the other cases.

The coordinate/momentum GH shifts, Eqs.~(\ref{eq7}) and (\ref{eq8}), and the Wigner time/frequency shifts, Eqs.~(\ref{eq9}) and (\ref{eq10}), can be associated with the real/imaginary parts of complex values $ i \partial \ln \{R,T\}/\partial k_y$ and $- i \partial \ln \{R,T\}/\partial \omega$. This duality and connections between the GH and Wigner effects were discussed in various contexts in Refs.~\cite{Kogelnik1974,Balcou1997,Asano2016} and is related to manifestations of complex quantum {\it weak values} \cite{Gotte2012,Toppel2013,Gorodetski2012,Asano2016,Jozsa2007,Dressel2014}.

\begin{figure}[!t]
\begin{center}
\includegraphics[width=0.95\linewidth]{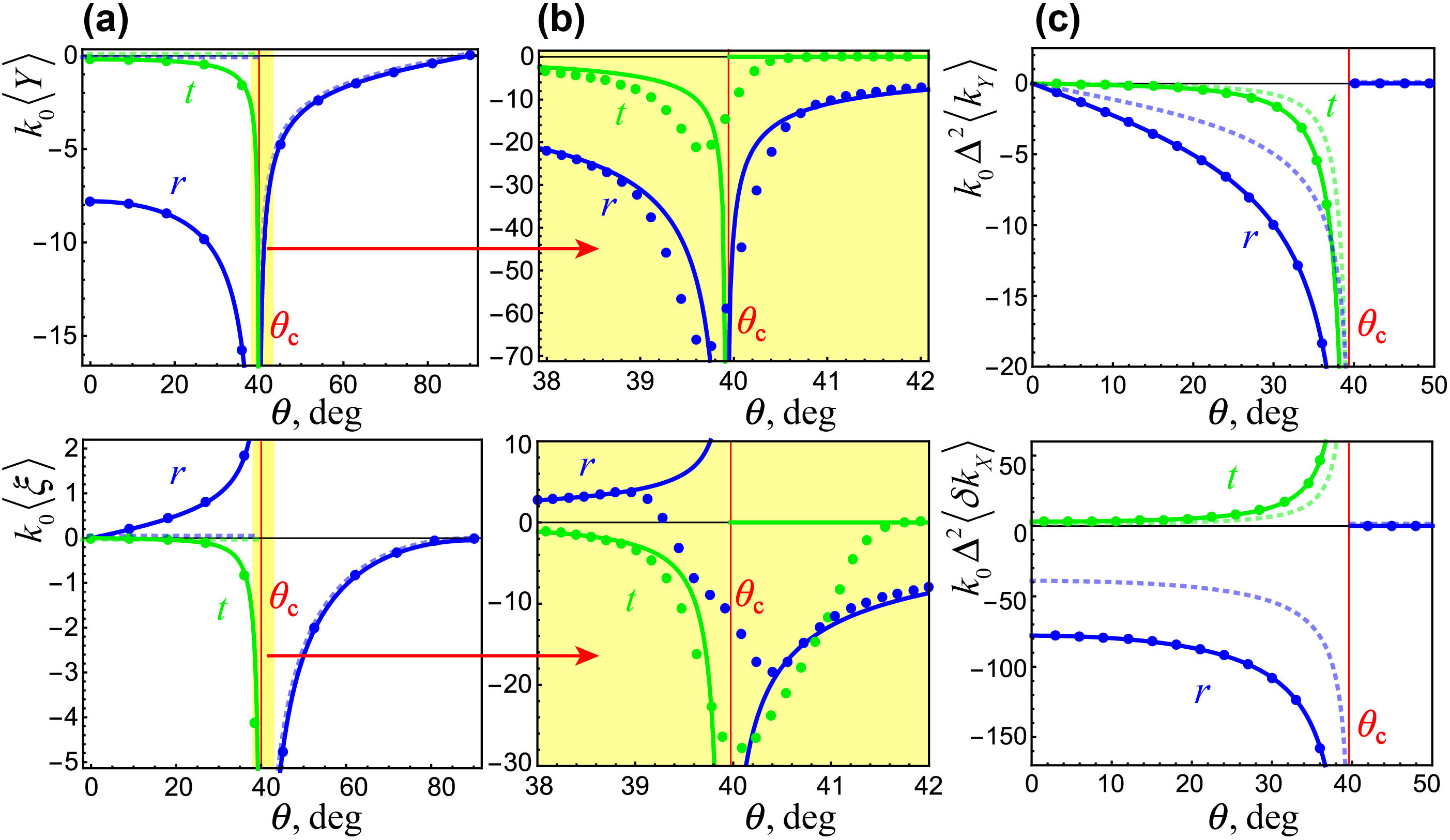}
\caption{
Theoretical (curves) and numerically calculated (symbols) GH shifts and Wigner time delays of the reflected ($r$) and transmitted ($t$) vortex wavepackets in the step-potential case (A) as functions of the angle of incidence $\theta$. 
The shifts are given by the sums of vortex-independent contributions, Eqs.~(\ref{eq7})--(\ref{eq10}), (shown by pale dotted curves) and the vortex-induced contributions, Eqs.~(\ref{eq14})--(\ref{eq16}). 
The linear (space-time) and angular (wavevector-energy) effects are shown in the panels (a) and (c), respectively.
Panels (b) provide zoom-in views of the resonant amplification of the linear shifts and time delays in the vicinity of the critical angle $\theta=\theta_c$. (The transmitted wavepacket exists for $\theta$ slightly exceeding $\theta_c$ because of the finite angular width $\delta\theta \sim (k_0\Delta)^{-1}$ of the plane-wave spectrum of the incident wavepacket.)
The parameters are: $E_0/V_0 =1.7$, $\ell=1$, $k_0\Delta = 628$, $\gamma = 0.4$. The time delays and energy shifts are presented via the corresponding longitudinal coordinate and wavevector shifts $\langle \xi^{r,t} \rangle = - v^{r,t}_g \langle \tau^{r,t} \rangle$ and $\langle \delta k^{r,t}_X \rangle \equiv \langle k^{r,t}_X - k^{r,t}_0 \rangle = (v^{r,t}_g)^{-1} \langle \epsilon^{r,t} \rangle$. Additional corrections to the shifts of transmitted wavepackets are considered in Section~4 and Fig.~\ref{Fig7}.
\label{Fig4}}
\end{center}
\end{figure}

The GH shifts and Wigner time delays, Eqs.~(\ref{eq7})--(\ref{eq10}) versus the angle of incidence $\theta$ in specific cases of barriers (A)--(C) are shown in Figures~\ref{Fig4}--\ref{Fig6} by dotted curves. 
Notice that the linear quantities $\langle Y \rangle$ and $\langle \tau \rangle$, Eqs.~(\ref{eq7}) and (\ref{eq9}), vanish in the case of partial reflection and transmission at the step potential (A), $\theta < \theta_c$, because the scattering coefficients (\ref{eq2}) are purely real in this case. The angular quantities $\langle k_Y \rangle$ and $\langle \epsilon \rangle$, Eqs.~(\ref{eq8}) and (\ref{eq10}), are generally nonzero in all cases (A)--(C). Importantly, the analytical expressions for the shifts and time delays can diverge at {\it zeros} of scattering coefficients: e.g., at the critical angle of incidence, $\theta_c$, in the case (A) and near zeros of the reflection coefficient (transmission resonances) in the case (C). Approximate formulas (\ref{eq7})--(\ref{eq10}) become inapplicable near such singularities, whereas the shifts/delays can exceed their typical values by several orders of magnitude \cite{Chan1985,Dasgupta2006,Merano2009,Berry2011,Asano2016,Hougne2021,Mazanov2021}. This amplification of otherwise small effects is a manifestation of quantum weak measurements and superweak values \cite{Gotte2013,Asano2016}.

\section{Scattering of 2D Laguerre-Gaussian-like vortices}


We are now in the position to consider a 2D vortex wavepacket. We model it with the spatiotemporal Laguerre-Gaussian solution \cite{Mazanov2021}, which adds an edge phase singularity (vortex) of the integer order $\ell$ in the center of the Gaussian wavepacket (\ref{eq5}) [see Fig.~\ref{Fig3}(b)]:
\begin{equation}
\tilde{\psi}(\mathbf{k}) \propto \left[ \gamma (k_X -k_0) +i \,{\rm sgn}(\ell) k_Y \right]^{|\ell |} 
\exp\! \left\{ -\frac{\Delta^{2}}{4}\left[\gamma^{2} (k_{X}-k_{0})^{2}+k_{Y}^{2}\right] \right\} .
\label{eq11} 
\end{equation}
In the paraxial approximation, neglecting diffraction effects, the Fourier integral of Eq.~(\ref{eq11}) yields the real-space form of the propagating 2D vortex \cite{Bliokh2012,Jhajj2016,Hancock2019,Chong2020,Bliokh2021,Zang2022,Junyi2022}:
\begin{equation}
\psi (\mathbf{r},t)  \propto {\left[ {{\gamma ^{ - 1}}\xi  + i\,{\rm sgn} ( \ell  ) Y} \right]^{\left| \ell  \right|}}\!
\exp\! \left[ { - \frac{\left( {{\gamma ^{ - 2}}{\xi ^2} + {Y^2}} \right)}{{{\Delta ^2}}} + i{k_0}X - i \omega_0 t } \right]\! .
\label{eq12} 
\end{equation}
%
For $\ell=0$, the wavepacket (\ref{eq11}) and (\ref{eq12}) becomes the Gaussian wavepacket (\ref{eq5}) and (\ref{eq6}). For $\ell \neq 0$, the wavefunction (\ref{eq12}) vanishes in the centre $(\xi,Y)={\bf 0}$, has the probability density $|\psi |^2$ in the form of an elliptical `doughnut' with the ratio of semiaxes $\gamma$, and a circulation of the probability current ${\rm Im}\! \left[ {{\psi ^*}{\bm \nabla}\psi } \right]$. The latter determines the normalized (per particle) OAM carried by the wavepacket and directed along the orthogonal $z$-axis \cite{Bliokh2012,Bliokh2021,Mazanov2021}:
\begin{equation}
\label{eq13}
\left\langle {{L_z}} \right\rangle = \frac{\iint{{\rm Im}\! \left[ {{\psi ^*}({\bf r} \times\! {\bm \nabla})_z \psi } \right]}\,dXdY}{\iint{{\psi ^*}\psi}\,dXdY}
= \frac{{\gamma  + {\gamma ^{ - 1}}}}{2}\,\ell \,.
\end{equation}
An example of the vortex wavepacket (\ref{eq11}) and (\ref{eq12}) with $\ell=1$ is shown in Fig.~\ref{Fig3}(b).

\begin{figure}[!t]
\begin{center}
\includegraphics[width=0.65\linewidth]{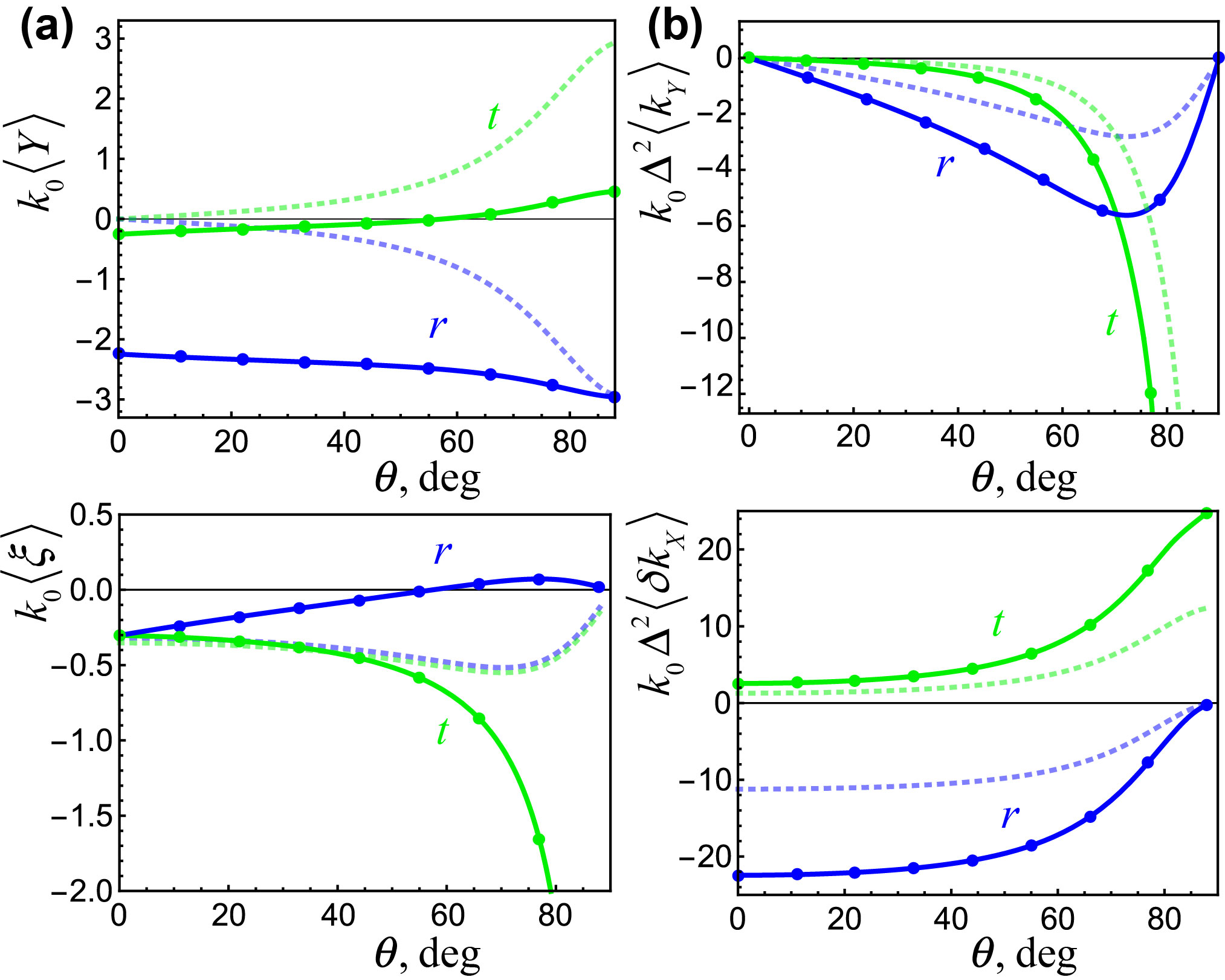}
\caption{Same as in Fig.~\ref{Fig4} but for the delta-function-barrier case (B) with $k_0/W_0 = 3$.
\label{Fig5}}
\end{center}
\end{figure}

When the vortex wavepacket (\ref{eq11}) and (\ref{eq12}) is scattered by a potential barrier, it experiences the GH shifts and Wigner time delays in the reflected and transmitted wavepackets. Here we examine how these effects are affected by the vortex in the incident packet. 
First, as was shown for optical vortex beams and spatiotemporal vortex pulses \cite{Bliokh2009,Merano2010,Mazanov2021}, the angular shifts are amplified with the factor of $(1+|\ell |)$ as compared to the shifts of Gaussian-type wavepackets, Eqs.~(\ref{eq8}) and (\ref{eq10}), so that the additional vortex-induced wavevector and energy shifts are:
\begin{equation}
\label{eq14}
\left\langle k_Y^{r,t} \right\rangle_\ell = |\ell | \left\langle k_Y^{r,t} \right\rangle_0 , \quad
\left\langle \epsilon^{r,t} \right\rangle_\ell = |\ell | \left\langle \epsilon^{r,t} \right\rangle_0 ,
\end{equation}

Second, the complex vortex structure $\left[ {{\gamma ^{ - 1}}\xi  + i\,{\rm sgn} ( \ell  ) Y} \right]^{\left| \ell  \right|}$ in the wavepacket (\ref{eq12}) produces a cross-coupling between the orthogonal Cartesian degrees of freedom. To describe it, we note that the reflected wavepacket will have a flipped vortex with the topological charge $-\ell$ and vortex structure $\left[ \gamma^{ - 1} \xi^r  - i\,{\rm sgn} ( \ell  ) Y^r \right]^{\left| \ell  \right|}$. Furthermore, in the potential-step case (A), the $X^t$ and $Y^t$ dimensions of the transmitted wavepacket are modified by the factors $k/k'$ and $\cos\theta' / \cos\theta$, respectively \cite{Bliokh2013,Mazanov2021}. This results in the transmitted vortex structure $\left[ (\gamma^t)^{ - 1} \xi^t  + i\,{\rm sgn} ( \ell  ) Y^t \right]^{\left| \ell  \right|}$, where $\gamma^t = \dfrac{k\cos\theta}{k^t\cos\theta^t}\, \gamma$. 

Next, we notice that the angular GH shifts $\left\langle k_Y^{r,t} \right\rangle_0$, Eq.~(\ref{eq8}), can be regarded as {\it imaginary} shifts in real space \cite{Bliokh2013,Bliokh2009,Mazanov2021,Aiello2008}: $\delta Y^r = - i \dfrac{\Delta^2}{2} \left\langle k_Y^r \right\rangle_0$ and $\delta Y^t = - i \dfrac{\Delta^2}{2}\dfrac{\cos^2\!\theta^t}{\cos^2\!\theta} \left\langle k_Y^t \right\rangle_0$. Substituting these imaginary shifts into the vortex forms of the corresponding scattered wavepackets, we find that they are equivalent to {\it real} $\ell$-dependent shifts in the {\it orthogonal} $\xi$-directions: $\left\langle \xi^r \right\rangle = - i \gamma\, \ell\, \delta Y^r$ and $\left\langle \xi^t \right\rangle = i \gamma^t\, \ell\, \delta Y^t$. In turn, these shifts are equivalent to the new vortex-induced time delays
\begin{equation}
\label{eq15}
\left\langle \tau^{r} \right\rangle_\ell = \ell \, \gamma \, \frac{\Delta^2}{2 v_g} \left\langle k_Y^r \right\rangle_0 , \quad
\left\langle \tau^{t} \right\rangle_\ell = - \ell \, \gamma \, \frac{\Delta^2}{2 v^t_g}\frac{k\cos\theta^t}{k^t\cos\theta} \left\langle k_Y^t \right\rangle_0 .
\end{equation}

In a similar manner, the angular Wigner shifts (\ref{eq10}), $\langle \delta k_X^{r,t} \rangle = (v^{r,t}_g)^{-1} \langle \epsilon^{r,t} \rangle$, can be regarded as imaginary shifts in real space: $\delta \xi^r = - i \dfrac{\gamma^2\Delta^2}{2} \left\langle \delta k_X^r \right\rangle_0$ and $\delta \xi^t = - i \dfrac{\gamma^2\Delta^2 k^2}{2\, {k^t}^2} \left\langle \delta k_X^t \right\rangle_0$.
Substituting these imaginary shifts into the vortex structures of the scattered wavepackets, we find that they are equivalent to real $\ell$-dependent shifts in the orthogonal $Y$-directions:
\begin{equation}
\label{eq16}
\left\langle Y^{r} \right\rangle_\ell = \ell \, \gamma \, \frac{\Delta^2}{2 v_g} \left\langle \epsilon^r \right\rangle_0 , \quad
\left\langle Y^{t} \right\rangle_\ell = - \ell \, \gamma \, \frac{\Delta^2}{2 v^t_g}\frac{k\cos\theta^t}{k^t\cos\theta} \left\langle \epsilon^t \right\rangle_0 .
\end{equation}
These are novel vortex-induced GH shifts. 

\begin{figure}[!t]
\begin{center}
\includegraphics[width=0.95\linewidth]{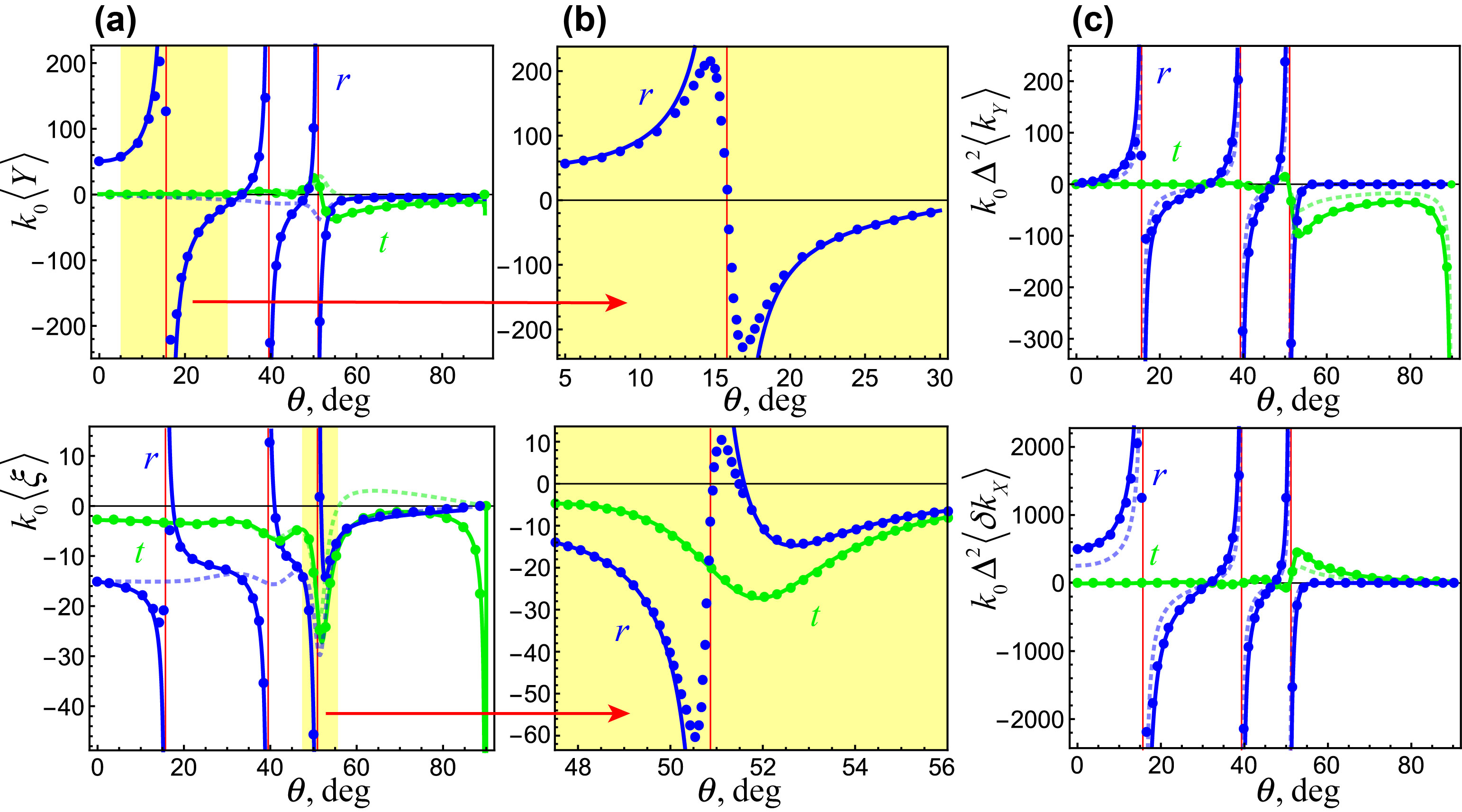}
\caption{Same as in Fig.~\ref{Fig4} but for the rectangular-barrier case (C) with $E_0/W_0 = 3$ and $k_0 a =5$. Panels (b) show zoom-in views of the resonant amplification of the GH shift and time delay of the reflected wavepacket near the transmission resonances (i.e., zeros of the reflection amplitude), cf. \cite{Asano2016,Hougne2021}.
\label{Fig6}}
\end{center}
\end{figure}

The resulting GH and Wigner shifts are given by sums of the corresponding Gaussian-packet shifts (\ref{eq7})--(\ref{eq10}) and $\ell$-dependent contributions (\ref{eq14})--(\ref{eq16}). Figures~\ref{Fig4}--\ref{Fig6} show these shifts for the vortex wavepacket with $\ell=1$ as functions of the angle of incidence $\theta$ in particular cases of barriers (A)--(C). In spite of the fact that the vortex-induced effects (\ref{eq14})--(\ref{eq16}) are expressed via the Gaussian-packet shifts (\ref{eq7})--(\ref{eq10}), the behaviour of the vortex-induced shifts and time differs dramatically. 
First, the vortex induced GH shifts $\langle Y^{r,t} \rangle$ and time delays $\langle \tau^{r,t} \rangle$ of the scattered vortex wavepackets are present even for purely real scattering amplitudes [e.g., for $\theta < \theta_c$ in the step-potential case (A)], when the corresponding Gaussian-packet effects vanish. 
Second, the GH shifts $\langle Y^{r,t} \rangle$ of vortex wavepackets are generally non-zero at normal incidence $\theta=0$. This is because the presence of a vortex breaks $y \to - y$ symmetry of the problem. 
Finally, the magnitude and sign of the vortex-induced shifts and time delays can be efficiently controlled by the vortex charge $\ell$, i.e., a parameter of the incident wavepacket. This can have implementations in vortex-induced transport phenomena, such as Hall and Magnus effects. We also emphasize the resonant amplification of the GH shifts and Wigner time delays in the vicinity of critical incidence $\theta=\theta_c$ in the case (A) and transmission resonances (i.e., zeros of the reflection amplitude) in the case (C). Approximate analytical formulas (\ref{eq7})--(\ref{eq10}) and (\ref{eq14})--(\ref{eq16}) diverge near such singularities, whereas the shifts/delays can exceed their typical values by several orders of magnitude \cite{Chan1985,Dasgupta2006,Merano2009,Berry2011,Gotte2013,Asano2016,Hougne2021,Mazanov2021}. 

\section{Numerical calculations and additional corrections}

\subsection{Equations for numerical calculations}

To check the above theoretical derivations, we performed numerical calculations of the reflection/transmission of a Laguerre-Gaussian vortex wavepacket by potential barriers (A)--(C) by applying exact boundary conditions and wavevector-dependent scattering amplitudes (\ref{eq2})--(\ref{eq4}) to each plane wave in the incident wavepacket spectrum (\ref{eq11}). This provides transformations of the wavefunctions in the momentum representation: $\tilde\psi^r ({\bf k}) = R({\bf k})\, \tilde\psi ({\bf k})$ and $\tilde\psi^t ({\bf k}) = T({\bf k})\, \tilde\psi ({\bf k})$. 
In the paraxial approximation, the scattering amplitudes can be expanded in the Taylor series near the central wavevector:
\begin{equation}
\label{eq17}
\tilde\psi^r ({\bf k}) \simeq \left[ R({\bf k}_0) + \dfrac{\partial R ({\bf k}_0)}{\partial {\bf k}_0} ({\bf k} - {\bf k}_0) \right]\! \tilde\psi ({\bf k}), ~~
\tilde\psi^t ({\bf k}) \simeq \left[ T({\bf k}_0) + \dfrac{\partial T ({\bf k}_0)}{\partial {\bf k}_0} ({\bf k} - {\bf k}_0) \right]\! \tilde\psi ({\bf k}).
\end{equation}

These equations should be supplied with the corresponding transformations of each wavevector 
 \cite{Bliokh2013,Gotte2012,Toppel2013,Mazanov2021}. Using the dispersion relations together with the conservation of the energy and $y$-components of the momentum (wavevector) at interfaces, we derive the following relations between the deflections of the wavevectors from their central values: 
\begin{eqnarray}
\label{eq18}
 \delta k^r_X  =  \delta k_X, \qquad k^r_Y = - k_Y,
  \nonumber\\
\delta k^t_X  \simeq  \frac{k_0}{k^t_0}\, \delta k_X + \frac{1}{2 k_0^t}\!\left( A\, \delta k_X^2 + B\, k_Y^2 + C\, \delta k_X k_Y \right)  ,   \nonumber\\
 k^t_Y  \simeq  \frac{\cos\theta}{\cos\theta^t}\, k_Y + D\, \delta k_X - \frac{\tan \theta^t}{2 k_0^t}\!\left( A\, \delta k_X^2 + B\, k_Y^2 + C\, \delta k_X k_Y \right).
\end{eqnarray}
Here $\delta k_X = k_X - k_0$, Eqs.~(\ref{eq18}) are derived in the second-order approximation in ${\bf k} - {\bf k}_0$, and we introduced auxiliary quantities
\begin{eqnarray}
\label{eq19}
A = \cot^2\!\theta\left(1-\frac{\cos^2\!\theta}{\cos^2\!\theta^t}\right), \quad
B = \left(1-\frac{\cos^2\!\theta}{\cos^2\!\theta^t}\right),
  \nonumber\\
C = - 2 \cot\theta \left(1-\frac{\cos^2\!\theta}{\cos^2\!\theta^t}\right), ~~
D = \frac{\cos\theta^t}{\sin\theta}\!\left(1-\frac{\cos^2\!\theta}{\cos^2\!\theta^t}\right).
\end{eqnarray}

The spatial and angular shifts are calculated as expectation values of the corresponding operators in the momentum representation: 
\begin{eqnarray}
\label{eq20}
\left\langle Y^{r,t} \right\rangle  = \frac{\left\langle \tilde{\psi}^{r,t} \right| i\dfrac{\partial}{\partial k_Y^{r,t}} \left| \tilde{\psi}^{r,t} \right\rangle}{\left\langle \tilde{\psi}^{r,t} \! \right.\left| \tilde{\psi}^{r,t} \right\rangle} , ~~
\left\langle \xi^{r,t} \right\rangle  = \frac{\left\langle \tilde{\psi}^{r,t} \right| i\dfrac{\partial}{\partial \delta k_X^{r,t}} \left| \tilde{\psi}^{r,t} \right\rangle}{\left\langle \tilde{\psi}^{r,t} \! \right.\left| \tilde{\psi}^{r,t} \right\rangle} ,~~\nonumber \\
\left\langle k_{Y}^{r,t} \right\rangle  = \frac{\left\langle \tilde{\psi}^{r,t} \right| k_{Y}^{r,t} \left| \tilde{\psi}^{r,t} \right\rangle}{\left\langle \tilde{\psi}^{r,t} \! \right.\left| \tilde{\psi}^{r,t} \right\rangle}, \quad
\left\langle \delta k_{X}^{r,t} \right\rangle  = \frac{\left\langle \tilde{\psi}^{r,t} \right| \delta k_{X}^{r,t} \left| \tilde{\psi}^{r,t} \right\rangle}{\left\langle \tilde{\psi}^{r,t} \! \right.\left| \tilde{\psi}^{r,t} \right\rangle},
\end{eqnarray}
where the inner product involves integration over the corresponding wavevector components $\left(\delta k^{r,t}_X,k^{r,t}_Y\right)$. Substituting the wavefunctions and wavevector components of the scattered wavepackets, Eqs.~(\ref{eq17})--(\ref{eq19}) into Eqs.~(\ref{eq20}), we obtain the formalism for calculations of all of the required shifts. In doing so, all quantities are expressed via the parameters of the incident wavepacket. In particular, the derivatives with respect to the wavevector components in the first two Eqs.~(\ref{eq20}) are expressed, using relations (\ref{eq18}) and (\ref{eq19}), as
\begin{eqnarray}
\label{eq21}
\frac{\partial}{\partial \delta k^r_X} =  \frac{\partial}{\partial \delta k_X} , \qquad  \frac{\partial}{\partial k^r_Y} = - \frac{\partial}{\partial k_Y},
  \nonumber\\
\frac{\partial}{\partial \delta k^t_X } \simeq  \frac{k^t_0}{k_0}\, \frac{\partial}{\partial \delta k_X } 
- \frac{\cos\theta^t}{\cos\theta}\,D\, \frac{\partial}{\partial k_Y } + \frac{\sin\theta^t}{k_0\cos\theta}
\left( A\, \delta k_X + C\, k_Y \right)  \frac{\partial}{\partial k_Y },   \nonumber\\
\frac{\partial}{\partial k^t_Y }  \simeq  \frac{\cos\theta^t}{\cos\theta}\, \frac{\partial}{\partial k_Y }  -
\frac{\cos\theta^t}{k_0\cos\theta} \left( B\, \delta k_Y + C\, \delta k_X \right)  \frac{\partial}{\partial \delta k_X }.
\end{eqnarray}

Importantly, for reflected wavepackets and transmitted wavepackets with $\theta^t = \theta$, i.e., in all cases apart from the refraction at the step potential, case (A), the coefficients (\ref{eq19}) vanish: $A=B=C=D=0$. In such case, Eqs.~(\ref{eq18}) and (\ref{eq21}) are simplified dramatically and become equivalent to standard equation for optical beams \cite{Bliokh2013,Gotte2012,Toppel2013,Mazanov2021}. 
The results of numerical calculations of Eqs.~(\ref{eq17})--(\ref{eq21}) with $A=B=C=D=0$ are depicted by symbols in Figs.~\ref{Fig4}--\ref{Fig6}. One can see that they perfectly agree with the analytical expressions (\ref{eq7})--(\ref{eq10}) and (\ref{eq14})--(\ref{eq16}) everywhere apart from the vicinity of resonant singularities. Analytical description of the resonant behaviour of wavepacket shifts, which regularizes divergencies of the Artmann and Wigner formulas, can be constructed within the nonlinear quantum-weak-weasurement approach \cite{Gorodetski2012,Asano2016}.

\subsection{Additional corrections for the refraction at a step potential}

The nontrivial case of the transmitted wavepacket at a step potential, case (A), when the coefficients (\ref{eq19}) do not vanish, requires additional considerations. We have shown that calculations without the $A$, $B$, $C$, $D$ terms in Eqs.~(\ref{eq18}) and (\ref{eq21}) yield the shifts described by equations (\ref{eq7})--(\ref{eq10}) and (\ref{eq14})--(\ref{eq16}). Hence, substituting the $A$, $B$, $C$, $D$ terms of Eqs.~(\ref{eq18}) and (\ref{eq21}) into Eqs.~(\ref{eq20}), we obtain corrections to the results of previous sections. For angular shifts, this yields 
\begin{equation}
\label{eq22}
 \langle k_Y^t \rangle = 
\langle k_Y^t \rangle_{0} + \langle k_Y^t \rangle_{\ell}+ 
v_g^{-1} D \left( \left\langle \epsilon^t \right\rangle_0 + \left\langle \epsilon^t \right\rangle_{\ell} \right) 
-  \frac{\left(1 + |\ell|\right)}{k_0 \Delta^2}  \frac{\sin^2\!\theta^t}{\sin\theta \cos\theta} 
\left( \gamma^{-2}\! A  + B \right), 
\end{equation}
\begin{equation}
\label{eq23}
\left\langle \epsilon^t \right\rangle = 
\left\langle \epsilon^t \right\rangle_0 + \left\langle \epsilon^t \right\rangle_{\ell} + 
\frac{v_g^t \left(1 + |\ell|\right)}{k_0 \Delta^2}  \frac{\sin^2\!\theta^t}{\sin\theta \cos\theta} 
\left( \gamma^{-2}\! A  + B \right).
\end{equation}
For linear shifts, we obtain 
\begin{equation}
\label{eq24}
 \langle Y^t \rangle = 
\langle Y^t \rangle_{0} + \langle Y^t \rangle_{\ell} 
-  \frac{\gamma \ell }{2 k_0}\,  \frac{\cos \theta^t}{\cos\theta}\, B \, , 
\end{equation}
\begin{equation}
\label{eq25}
\left\langle \tau^t \right\rangle = 
\left\langle \tau^t \right\rangle_0 + \left\langle \tau^t \right\rangle_{\ell} + v_g^{-1} D \left( \left\langle Y^t \right\rangle_0 + \left\langle Y^t \right\rangle_{\ell} \right)  +
\frac{\ell }{2 \gamma v_g k_0}\,  \frac{\sin \theta^t}{\cos\theta}\, A \, .
\end{equation}
Remarkably the $A$, $B$, $D$ corrections in Eqs.~(\ref{eq22})--(\ref{eq25}) affect even the shifts of Gaussian wavepackets, $\ell=0$. Thus, that the standard Artmann and Wigner formulae (\ref{eq7})--(\ref{eq10}) become inaccurate in the case of the refraction at a step potential.  

\begin{figure}[!t]
\begin{center}
\includegraphics[width=0.95\linewidth]{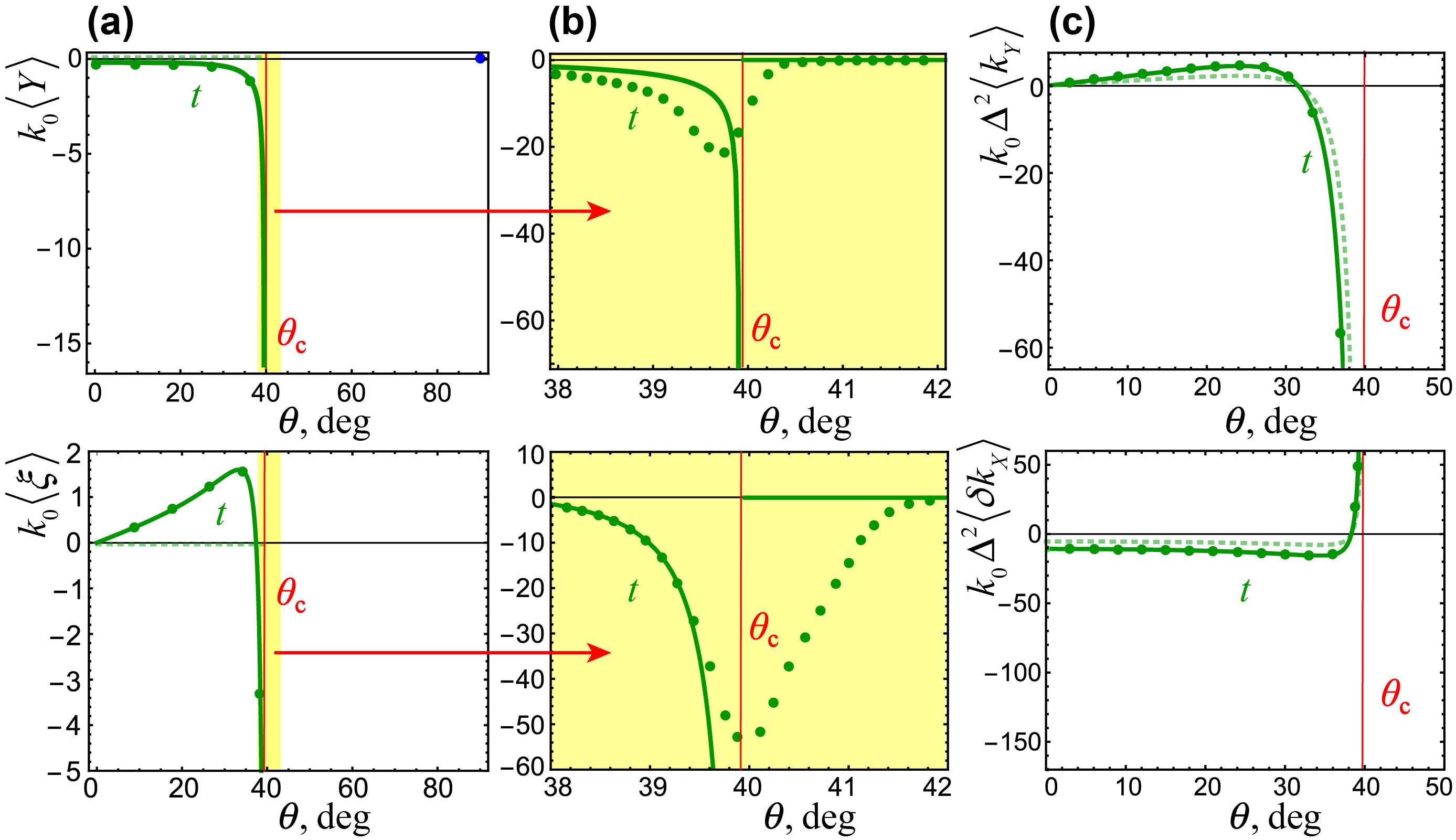}
\caption{Same as in Fig.~\ref{Fig4} for transmitted wavepackets but with the $A$, $B$, $C$, $D$ corrections in Eqs.~(\ref{eq17})--(\ref{eq25}).
\label{Fig7}}
\end{center}
\end{figure}

Figure~\ref{Fig7} shows the linear and angular shifts of transmitted wavepackets at a step potential, both analytical expressions (\ref{eq22})--(\ref{eq25}) and numerical calculations using Eqs.~(\ref{eq17})--(\ref{eq21}). By comparing these plots with Fig.~\ref{Fig4} one can see that the $A$, $B$, $C$, $D$ corrections can considerably modify the values of the shifts and even change their signs. 

\section{Concluding remarks}

We have described reflection and transmission of a localized 2D quantum vortex wavepacket at a planar potential barrier. We considered elliptical Laguerre-Gaussian-type wavepackets and step-like, delta-function, and rectangular potentials. Employing the analogy with the previously analysed reflection/refraction of optical vortex beams and spatiotemporal pulses, we have derived analytical expressions for the GH shifts and Wigner time delays of the reflected and transmitted wavepackets. In doing so, both `linear' (space-time) and `angular' (wavevector-energy) shifts were calculated. (The angular shifts have been mostly ignored so far in quantum problems.) 

Importantly, the presence of a vortex dramatically modifies these shifts, previously known only for Gaussian-type wavepackets.
First, the vortex-modified linear shifts and time delays appear even for purely real scattering coefficients, where the standard Artmann and Wigner expressions vanish. 
Second, the GH shifts of vortex wavepackets are generally non-zero even at normal incidence.
Finally, the magnitudes and signs of all the vortex-induced shifts can be controlled by the topological charge of the vortex, $\ell$. Furthermore, we have shown that the shifts and time delays can be resonantly enhanced by several orders of magnitude near the critical angle of incidence for a step potential and near zeros of the reflection amplitude (transmission resonances) for a rectangular barrier.
 
In addition to the analytical expressions, we have performed numerical calculations of the GH and Wigner shifts using the Fourier plane-wave expansions of the incident and scattered wavepackets. 
One can expect that the new shifts and time delays described in our work can manifest themselves in vortex-dependent transport phenomena in 2D quantum systems, including superfluids, quantum-Hall systems, 2D electron gas, ferromagnets, etc. 
While here we considered scattering of a vortex wavepacket by a planar scalar potential, it is worth mentioning that another lateral shift phenomenon appears upon scattering of a Gaussian wavepacket by a vortex Aharonov-Bohm vector-potential \cite{Shelankov1998,Berry1999}. This phenomenon also owes its origin to the fine interference of plane waves in the scattered wavepacket. 

\section*{Acknowledgements}
We are grateful to 
Professor M. V. Berry 
for providing an ongoing inspiration to play with waves, phases, currents, and singularities. 
This work is dedicated to him on the occasion of his 80th birthday.

\section*{References}

\bibliography{References}

\providecommand{\newblock}{}
\begin{thebibliography}{10}
\expandafter\ifx\csname url\endcsname\relax
  \def\url#1{{\tt #1}}\fi
\expandafter\ifx\csname urlprefix\endcsname\relax\def\urlprefix{URL }\fi
\providecommand{\eprint}[2][]{\url{#2}}

\bibitem{Nye1974}
Nye J~F and Berry M~V 1974 \titlecap{Dislocations in wave trains} {\em Proc. R.
  Soc. Lond. Ser. A\/} {\bf 336} 165--190

\bibitem{Berry1981}
Berry M~V 1981 Singularities in Waves {\em {Les Houches Lecture Series Session
  XXXV}\/} ed Balian R, Kleman M and Poirier J~P (North-Holland: Amsterdam) pp
  453--543

\bibitem{Rubinsztein2016}
Rubinsztein-Dunlop H, Forbes A, Berry M~V, Dennis M~R, Andrews D~L, Mansuripur
  M, Denz C, Alpmann C, Banzer P, Bauer T, Karimi E, Marrucci L, Padgett M,
  Ritsch-Marte M, Litchinitser N~M, Bigelow N~P, Rosales-Guzm{\'{a}}n C,
  Belmonte A, Torres J~P, Neely T~W, Baker M, Gordon R, Stilgoe A~B, Romero J,
  White A~G, Fickler R, Willner A~E, Xie G, McMorran B and Weiner A~M 2016
  \titlecap{Roadmap on structured light} {\em J. Opt.\/} {\bf 19} 013001

\bibitem{Aharonov1988}
Aharonov Y, Albert D~Z and Vaidman L 1988 \titlecap{How the result of a
  measurement of a component of the spin of a spin-1/2 particle can turn out to
  be 100} {\em Phys. Rev. Lett.\/} {\bf 60} 1351--1354

\bibitem{Berry2019}
Berry M, Zheludev N, Aharonov Y, Colombo F, Sabadini I, Struppa D~C, Tollaksen
  J, Rogers E~T~F, Qin F, Hong M, Luo X, Remez R, Arie A, G{\"o}tte J~B, Dennis
  M~R, Wong A~M~H, Eleftheriades G~V, Eliezer Y, Bahabad A, Chen G, Wen Z,
  Liang G, Hao C, Qiu C~W, Kempf A, Katzav E and Schwartz M 2019
  \titlecap{Roadmap on superoscillations} {\em J. Opt.\/} {\bf 21} 053002

\bibitem{Dressel2014}
Dressel J, Malik M, Miatto F~M, Jordan A~N and Boyd R~W 2014
  \titlecap{Colloquium: Understanding quantum weak values: Basics and
  applications} {\em Rev. Mod. Phys.\/} {\bf 86} 307--316

\bibitem{Chiao1997}
Chiao R~Y and Steinberg A~M 1997 \titlecap{Tunneling times and superluminality}
  {\em Prog. Opt.\/} {\bf 37} 345--405

\bibitem{Carvalho2002}
de~Carvalho C~A~A and Nussenzveig H~M 2002 \titlecap{Time delay} {\em Phys.
  Rep.\/} {\bf 364} 83--174

\bibitem{Winful2006}
Winful H~G 2006 \titlecap{Tunneling time, the Hartman effect, and
  superluminality: A proposed resolution of an old paradox} {\em Phys. Rep.\/}
  {\bf 436} 1--69

\bibitem{Bliokh2013}
Bliokh K~Y and Aiello A 2013 \titlecap{Goos-H\"{a}nchen and Imbert-Fedorov beam
  shifts: an overview} {\em J. Opt.\/} {\bf 15} 014001

\bibitem{Gotte2012}
G\"otte J~B and Dennis M~R 2012 \titlecap{Generalized shifts and weak values
  for polarization components of reflected light beams} {\em New J. Phys.\/}
  {\bf 14} 073016

\bibitem{Toppel2013}
T\"{o}ppel F, Ornigotti M and Aiello A 2013 \titlecap{Goos-H\"{a}nchen and
  Imbert-Fedorov shifts from a quantum-mechanical perspective} {\em New J.
  Phys.\/} {\bf 15} 113059

\bibitem{Bliokh2015NP}
Bliokh K~Y, Rodr{\'\i}guez-Fortu{\~n}o F~J, Nori F and Zayats A~V 2015
  \titlecap{Spin-orbit interactions of light} {\em Nat. Photon.\/} {\bf 9}
  796--808

\bibitem{Xiao2017}
Xiao S, Wang J, Liu F, Zhang S, Yin X and Li J 2017 \titlecap{Spin-dependent
  optics with metasurfaces} {\em Nanophotonics\/} {\bf 6} 215--234

\bibitem{Hosten2008}
Hosten O and Kwiat P 2008 \titlecap{Observation of the Spin Hall Effect of
  Light via Weak Measurements} {\em Science\/} {\bf 319} 787--790

\bibitem{Berry2009}
Berry M~V 2009 \titlecap{Optical currents} {\em J. Opt. A: Pure Appl. Opt.\/}
  {\bf 11} 094001

\bibitem{Bliokh2013NJP_II}
Bliokh K~Y, Bekshaev A~Y, Kofman A~G and Nori F 2013 \titlecap{Photon
  trajectories, anomalous velocities and weak measurements: a classical
  interpretation} {\em New J. Phys.\/} {\bf 15} 073022

\bibitem{Asano2016}
Asano M, Bliokh K~Y, Bliokh Y~P, Kofman A~G, Ikuta R, Yamamoto T, Kivshar Y~S,
  Yang L, \"{O}zdemir N~I~S~K and Nori F 2016 \titlecap{Anomalous time delays
  and quantum weak measurements in optical micro-resonators} {\em Nat.
  Commun.\/} {\bf 7} 13488

\bibitem{Dragoman}
Dragoman D and Dragoman M 2004 {\em Quantum-Classical Analogies\/} (Springer)

\bibitem{Wigner1954}
Wigner E~P 1954 \titlecap{Lower Limit for the Energy Derivative of the
  Scattering Phase Shift} {\em Phys. Rev.\/} {\bf 98} 145--147

\bibitem{Goos1947}
Goos F and H\"{a}nchen H 1947 \titlecap{Ein neuer und fundamentaler Versuch zur
  Totalreflexion} {\em Ann. Phys.\/} {\bf 1} 333--346

\bibitem{Artmann1948}
Artmann K 1948 \titlecap{Berechnung der Seitenversetzung des totalreflektierten
  Strahles} {\em Ann. Phys.\/} {\bf 2} 87--102

\bibitem{Fedoseyev2001}
Fedoseyev V~G 2001 \titlecap{Spin-independent transverse shift of the centre of
  gravity of a reflected and of a refracted light beam} {\em Opt. Commun.\/}
  {\bf 193} 9--18

\bibitem{Dasgupta2006}
Dasgupta R and Gupta P~K 2006 \titlecap{Experimental observation of
  spin-independent transverse shift of the centre of gravity of a reflected
  Laguerre-Gaussian light beam} {\em Opt. Commun.\/} {\bf 257} 91--96

\bibitem{Fedoseyev2008}
Fedoseyev V~G 2008 \titlecap{Transformation of the orbital angular momentum at
  the reflection and transmission of a light beam on a plane interface} {\em
  {J. Phys. A: Math. Theor.}\/} {\bf 41} 505202

\bibitem{Bliokh2009}
Bliokh K~Y, Shadrivov I~V and Kivshar Y~S 2009 \titlecap{Goos-H\"{a}nchen and
  Imbert-Fedorov shifts of polarized vortex beams} {\em Opt. Lett.\/} {\bf 34}
  389--391

\bibitem{Merano2010}
Merano M, Hermosa N, Woerdman J~P and Aiello A 2010 \titlecap{How orbital
  angular momentum affects beam shifts in optical reflection} {\em Phys. Rev.
  A\/} {\bf 82} 023817

\bibitem{Dennis2012}
Dennis M~R and G\"{o}tte J~B 2012 \titlecap{Topological Aberration of Optical
  Vortex Beams: Determining Dielectric Interfaces by Optical Singularity
  Shifts} {\em Phys. Rev. Lett.\/} {\bf 109} 183903

\bibitem{Bliokh2012}
Bliokh K~Y and Nori F 2012 \titlecap{Spatiotemporal vortex beams and angular
  momentum} {\em Phys. Rev. A\/} {\bf 86} 033824

\bibitem{Jhajj2016}
Jhajj N, Larkin I, Rosenthal E~W, Zahedpour S, Wahlstrand J~K and Milchberg H~M
  2016 \titlecap{Spatiotemporal Optical Vortices} {\em Phys. Rev. X\/} {\bf 6}
  031037

\bibitem{Hancock2019}
Hancock S~W, Zahedpour S, Goffin A and Milchberg H~M 2019 \titlecap{Free-space
  propagation of spatiotemporal optical vortices} {\em Optica\/} {\bf 6} 1547

\bibitem{Chong2020}
Chong A, Wan C, Chen J and Zhan Q 2020 \titlecap{Generation of spatiotemporal
  optical vortices with controllable transverse orbital angular momentum} {\em
  Nat. Photon.\/} {\bf 14} 350

\bibitem{Bliokh2021}
Bliokh K~Y 2021 \titlecap{Spatiotemporal Vortex Pulses: Angular Momenta and
  Spin-Orbit Interaction} {\em Phys. Rev. Lett.\/} {\bf 126} 243601

\bibitem{Zang2022}
Zang Y, Mirando A and Chong A 2022 \titlecap{Spatiotemporal optical vortices
  with arbitrary orbital angular momentum orientation by astigmatic mode
  converters} {\em Nanophotonics\/} {\bf 11} 745--752

\bibitem{Junyi2022}
Huang J, Zhang J, Zhu T and Ruan Z 2022 \titlecap{Spatiotemporal
  Differentiators Generating Optical Vortices with Transverse Orbital Angular
  Momentum and Detecting Sharp Change of Pulse Envelope} {\em Laser Photonics
  Rev.\/}  2100357

\bibitem{Mazanov2021}
Mazanov M, Sugic D, Alonso M~A, Nori F and Bliokh K~Y 2021 \titlecap{Transverse
  shifts and time delays of spatiotemporal vortex pulses reflected and
  refracted at a planar interface} {\em Nanophotonics\/}  20210294

\bibitem{Huebener}
Huebener R~P, Schopohl N and Volovik G~E (eds) 2002 {\em Vortices in
  Unconventional Superconductors and Superfluids\/} (Springer)

\bibitem{Fetter2009}
Fetter A~L 2009 \titlecap{Rotating trapped Bose-Einstein condensates} {\em Rev.
  Mod. Phys.\/} {\bf 81} 647--691

\bibitem{Nagaosa2013}
Nagaosa N and Tokura Y 2013 \titlecap{Topological properties and dynamics of
  magnetic skyrmions} {\em Nature Nanotechnology\/} {\bf 8} 899--911

\bibitem{Thouless1996}
Thouless D~J, Ao P and Niu Q 1996 \titlecap{Transverse Force on a Quantized
  Vortex in a Superfluid} {\em Phys. Rev. Lett.\/} {\bf 76} 3758--3761

\bibitem{Stone1996}
Stone M 1996 \titlecap{Magnus force on skyrmions in ferromagnets and quantum
  Hall systems} {\em Phys. Rev. B\/} {\bf 53} 16573--16578

\bibitem{Thaller_book}
Thaller B 2000 {\em Visual Quantum Mechanics\/} (Springer, Berlin)

\bibitem{Chan1985}
Chan C~C and Tamir T 1985 \titlecap{Angular shift of a Gaussian beam reflected
  near the Brewster angle} {\em Opt. Lett.\/} {\bf 10} 378--380

\bibitem{Aiello2008}
Aiello A and Woerdman J~P 2008 \titlecap{Role of beam propagation in
  Goos-H\"{a}nchen and Imbert-Fedorov shifts} {\em Opt. Lett.\/} {\bf 33}
  1437--1439

\bibitem{Merano2009}
Merano M, Aiello A, van Exter M~P and Woerdman J~P 2009 \titlecap{Observing
  angular deviations in the specular reflection of a light beam} {\em Nat.
  Photon.\/} {\bf 3} 337--340

\bibitem{Beenakker2009}
Beenakker C~W~J, Sepkhanov R~A, Akhmerov A~R and Tworzyd\l{}o J 2009
  \titlecap{Quantum Goos-H\"anchen Effect in Graphene} {\em Phys. Rev. Lett.\/}
  {\bf 102} 146804

\bibitem{Haan2010}
de~Haan V~O, Plomp J, Rekveldt T~M, Kraan W~H, van Well A~A, Dalgliesh R~M and
  Langridge S 2010 \titlecap{Observation of the Goos-H\"anchen Shift with
  Neutrons} {\em Phys. Rev. Lett.\/} {\bf 104} 010401

\bibitem{Wu2011}
Wu Z, Zhai F, Peeters F~M, Xu H~Q and Chang K 2011 \titlecap{Valley-Dependent
  Brewster Angles and Goos-H\"anchen Effect in Strained Graphene} {\em Phys.
  Rev. Lett.\/} {\bf 106} 176802

\bibitem{Chen2013}
Chen X, Lu X~J, Ban Y and Li C~F 2013 \titlecap{Electronic analogy of the
  Goos{\textendash}H{\"a}nchen effect: a review} {\em J. Opt.\/} {\bf 15}
  033001

\bibitem{Kogelnik1974}
Kogelnik H and Weber H~P 1974 \titlecap{Rays, stored energy, and power flow in
  dielectric waveguides$\ast$} {\em J. Opt. Soc. Am.\/} {\bf 64} 174--185

\bibitem{Balcou1997}
Balcou P and Dutriaux L 1997 \titlecap{Dual Optical Tunneling Times in
  Frustrated Total Internal Reflection} {\em Phys. Rev. Lett.\/} {\bf 78}
  851--854

\bibitem{Gorodetski2012}
Gorodetski Y, Bliokh K~Y, Stein B, Genet C, Shitrit N, Kleiner V, Hasman E and
  Ebbesen T~W 2012 \titlecap{Weak Measurements of Light Chirality with a
  Plasmonic Slit} {\em Phys. Rev. Lett.\/} {\bf 109} 013901

\bibitem{Jozsa2007}
Jozsa R 2007 \titlecap{Complex weak values in quantum measurement} {\em Phys.
  Rev. A\/} {\bf 76} 044103

\bibitem{Berry2011}
Berry M~V 2011 \titlecap{Lateral and transverse shifts in reflected dipole
  radiation} {\em Proc. R. Soc. A\/} {\bf 467} 2500--2519

\bibitem{Hougne2021}
del Hougne P, Yeo K~B, Besnier P and Davy M 2021 \titlecap{On-Demand Coherent
  Perfect Absorption in Complex Scattering Systems: Time Delay Divergence and
  Enhanced Sensitivity to Perturbations} {\em Laser Photonics Rev.\/} {\bf 15}
  2000471

\bibitem{Gotte2013}
G\"{o}tte J~B and Dennis M~R 2013 \titlecap{Limits to superweak amplification
  of beam shifts} {\em Opt. Lett.\/} {\bf 38} 2295--2297

\bibitem{Shelankov1998}
Shelankov A~L 1998 \titlecap{Magnetic force exerted by the Aharonov-Bohm line}
  {\em Europhys. Lett. ({EPL})\/} {\bf 43} 623--628

\bibitem{Berry1999}
Berry M~V 1999 \titlecap{Aharonov-Bohm beam deflection:
  Shelankov{\textquotesingle}s formula, exact solution, asymptotics and an
  optical analogue} {\em J. Phys. A: Math. Gen.\/} {\bf 32} 5627--5641

\end{thebibliography}

\end{document}